%% file: main.tex
\documentclass[smallextended]{svjour3}       

\smartqed  

\usepackage[utf8]{inputenc}
\usepackage{graphicx}
\usepackage{authblk}
\usepackage[caption=false]{subfig}
\usepackage{soul,color}
\usepackage[sort&compress,numbers]{natbib}
\usepackage{xspace}
\usepackage{amsfonts}
\usepackage{amsmath}
\usepackage{lipsum}
\usepackage{booktabs}
\usepackage{comment}
\usepackage[linesnumbered,algoruled,boxed,lined,vlined]{algorithm2e}
\usepackage{todonotes}

\usepackage{layouts}

\input{s00-defs.tex}

\journalname{The Journal of Supercomputing}

\begin{document}

\title{Leveraging knowledge-as-a-service (KaaS) for QoS-aware resource management
  in multi-user video transcoding%
  \thanks{This work has been partially supported by the EU (FEDER), the Spanish
    MINECO and the CM under grants S2018/TCS-4423, TIN 2015-65277-R and
    RTI2018-093684-B-I00 and the Spanish MECD under grant FPU15/02050.}%
}

\titlerunning{Leveraging Knowledge-as-a-service (KaaS) for resource
management \ldots}

\author{%
  Luis Costero  \and
  Francisco D. Igual \and
  Katzalin Olcoz \and
  Francisco Tirado
}
\authorrunning{L. Costero, F. D. Igual, K. Olcoz, P. Tirado} 

\institute{Departamento de Arquitectura de Computadores y Automática \at
  Universidad Computense de Madrid\\
  \email{\{lcostero,figual,katzalin,ptirado\}@ucm.es}
}

\date{Received: date / Accepted: date}
\maketitle

\input{abstract.tex}

\input{s1-intro.tex}

\input{s2-policies_design.tex}

\input{s3-policies_results.tex}
\input{s4-heuristic_design.tex}

\input{s5-heuristic_results.tex}
\input{s6-related_work.tex}
\input{s7-conclusions.tex}

\bibliographystyle{spbasic}      

\end{document}

%% file: s00-defs.tex
\newcommand{\newtext}[1]{#1}
\newenvironment{newtext_env}{}{}

\newcommand{\mdp}{MDP\xspace}
\newcommand{\rl}{RL\xspace}

\newcommand{\regular}{{\em regular}\xspace}
\newcommand{\premium}{{\em premium}\xspace}

\newcommand{\onepol}{\mbox{{\sc 1Pol\newtext{-Mamut}}}\xspace}
\newcommand{\twopol}{{\sc 2Pol}\xspace}
\newcommand{\heupol}{{\sc 3TierPol}\xspace}

\newcommand{\policyR}{$\pi^{R}$\xspace}
\newcommand{\policyP}{$\pi^{P}$\xspace}
\newcommand{\policyRm}{$\pi_{R}$\xspace}
\newcommand{\policyPm}{$\pi_{P}$\xspace}

\newcommand{\PSNR}{{\sc psnr}\xspace}

\newcommand{\RpsnrH}{$R_{PSNR-H}$\xspace}
\newcommand{\RpsnrM}{$R_{PSNR-M}$\xspace}
\newcommand{\RpsnrL}{$R_{PSNR-L}$\xspace}
\newcommand{\Rfps}{$R_{FPS}$\xspace}
\newcommand{\Rpower}{$R_{POWER}$\xspace}

%% file: abstract.tex
\begin{abstract}

The coexistence of parallel applications in shared computing nodes, each one featuring different
Quality of Service (QoS) requirements, carries out new challenges to improve resource occupation while keeping acceptable rates in terms of QoS.
As more application-specific and system-wide metrics are included as QoS dimensions,
or under situations in which resource-usage limits are strict, 
building and serving the most appropriate set of actions (application
control knobs and system resource assignment)
to concurrent applications in an automatic and optimal fashion becomes mandatory.
In this paper, we propose strategies to build and serve this type of knowledge to concurrent applications 
by leveraging Reinforcement Learning techniques. 
Taking multi-user video transcoding as a driving example, our experimental results reveal an
excellent adaptation of resource and knob management to heterogeneous QoS requests, and
increases in the amount of concurrently served users up to $1.24\times$ compared with
alternative approaches considering homogeneous QoS requests.

\keywords{Resource management \and Heterogeneous Quality of Service \and Reinforcement Learning \and Multi-core architectures \and HEVC video transcoding}
\end{abstract}

%% file: s1-intro.tex
\section{Introduction and motivation}
\label{sec:intro}

The integration of intelligent policies for resource management and application tuning 
in shared computing systems is becoming a field of paramount interest to efficiently 
exploit the potential of the underlying architectures without human intervention. In
situations where external limitations in terms of Quality of Service (QoS), tight
per-application SLA (Service Level Agreements) or energy consumption are imposed, 
the development and application 
of such policies becomes a hurdle difficult to be automatically addressed~\cite{Wang2018}. 



%

Two of the main properties of any generic autonomous system, including resource managers
in the fields of Cloud Computing or HPC,
are {\em self-configuration} (ability to adapt to environmental changes) and {\em self-optimization} (capability to improve performance and reduce overloading or underloading the underlying resources)~\cite{Singh2015}. 
Resource managers can be actually considered in terms of the IBM Autonomic Model~\cite{ibm2005architectural},
which encompasses four main generic steps. 
This sequence of steps is repeated in a control loop, that
usually features sensoring and acting capabilities towards the underlying architecture or application. 
These actions are commonly cast in terms of selecting 
values for architectural knobs (e.g. core frequency) or application-specific knobs. 
In addition, these cyclic steps orbit around the existence of a shared Knowledge Base ({\sc KB})
storing rules that, properly orchestrated, can fulfill the requirements and restrictions
imposed without further human intervention.

The development of the {\sc KB}, however, can become a daunting task when the amount of architectural and
application-level knobs increase and their interplay is nontrivial. 
In stochastic environments such as shared
nodes in cloud deployments, in which the application of a given rule does not always yield the same result in terms 
of performance and/or energy consumption, the creation, maintenance and effective application of the knowledge of the {\sc KB} is even a more complex task. The challenge is harder in scenarios
in which the request arrival rate and its distribution are unknown, 
or when the throughput or quality attained are content-dependent, and hence unpredictable.

Reinforcement Learning (RL)~\cite{RL} is a field of Artificial Intelligence that 
has shown to be effective in problems that feature complex and large state spaces, dynamic environments and without any pre-established knowledge.
%
In this paper, we leverage RL to build, maintain and enrich the Knowledge Base of a centralized 
resource manager in shared servers in order to attain automatic and efficient resource management and assignation 
for multiple concurrent applications exhibiting {\em heterogeneous} QoS demands. 
%

This idea, that can be considered as 
{\em Knowledge-as-a-service} (KaaS), is illustrated in this paper by means of 
a  practical yet illustrative example:  QoS-aware multi-user video transcoding.
In this scenario, multiple users demand the execution of multiple
concurrent transcoding processes, each one with different QoS requirements. 
We consider two types of users with different QoS requirements
(regular and premium users), and optional resource-minimization strategies.

Resource allocation and management in shared computing environments has been
thoroughly studied in the past from different perspectives.~\cite{Sembiring2013}
studied the differences between static and dynamic resource allocation for media
processing. 
Dynamic resource provisioning has also been tackled in 
terms of multiple priorities for heterogeneous QoS requirements in~\cite{Gao2019, Gao2016}.
Based on simulation,~\cite{Farhad2016} studies policies for allocating and deallocating
virtual machines to attend multiple QoS requirements by concurrent applications. Typically,
delay time is the metric considered in these works as the only QoS target, ignoring
other application-specific quality metrics and architecture-specific considerations (e.g.
energy consumption).
Many of these works propose predictions or model-based projections for
resource usage and QoS violations to apply heuristics and policies for resource
management. Reinforcement Learning has been shown to be a useful technique to develop
self-adaptive policies for multi-QoS scenarios, with model-based approaches~\cite{Ho2015}
or model-free approaches~\cite{Iranfar18}.

Our work introduces the following novelties compared with previous works:

\begin{itemize}

\item We integrate RL techniques into a centralized
resource manager, and demonstrate its effectiveness to attend 
multi-user environments with {\em heterogeneous QoS requirements} within
a single compute node.

\item We formalize a methodology to build different policies based on RL, 
that ultimately make it possible to learn different policies serving different
QoS requirements in a feasible amount of time.

\item We propose strategies to boost learning time based on the construction of the
transitions table of the system, used to train the RL system.


\item We demonstrate the effectiveness of our proposal on a real-time HEVC encoding application. We deal with extended sets of both application- and system-level knobs. For
multi-user video transcoding, we consider quality, throughput and energy
consumption as target metrics to be simultaneously improved and/or limited.

\item We compare the learning time benefits of our approach against other state-of-the-art RL approaches.

\item 
We propose heuristics to serve different knowledge to individual users in real-time on a shared computing node.

\item Reported results are based on an actual multi-core architecture, demonstrating the
ability of RL to deal with stochastic (noisy) environments. 

\end{itemize}

The rest of the paper is structured as follows.
Section~\ref{sec:policies_design} proposes efficient mechanisms to build
ad-hoc policies leveraging Reinforcement Learning. 
Section~\ref{sec:policies_results} motivates the use of multi-user
video transcoding as an illustrative example of such techniques and reports the attained results for different
policies designed to adapt multi-user video transcoding to different levels of
QoS and resource usage.
Section~\ref{sec:heuristic_design} introduces heuristics to leverage the existence
of different policies in the KB, and to apply them in scenarios with heterogeneous
QoS requirements.
In Section~\ref{sec:heuristic_results}, we report the observed results and benefits
of such heuristics.
Section~\ref{sec:conclusions} closes the paper with some final remarks.



%% file: s2-policies_design.tex
\section{Policy design}
\label{sec:policies_design}


A {\it Markov Decision Process} (\mdp) is a framework for modelling decision making problems in stochastic environments. A MDP=$(\mathcal{S},\mathcal{A},\mathcal{P},\mathcal{R})$ is defined by a finite set of states the system can be at each moment of the execution, $\mathcal{S}$, a finite set of actions, $\mathcal{A}$, that can be applied to the system and can produce a change in the state the system is, a set of probability functions $\mathcal{P} =  \{ P_a(s,s'):\mathcal{S}\times\mathcal{S}\times\mathcal{A}\rightarrow[0,1]\}$ which determine the probability of moving from one state ($s$) to another ($s'$) after taking an action ($a$), and a set of reward functions $\mathcal{R}=\{R_a(s,s'):\mathcal{S}\times\mathcal{S}\times\mathcal{A}\rightarrow\mathbb{R}\}$ which define how good or bad was the transition $s\rightarrow s'$ due to action $a$. The final goal of a \mdp is to find a policy $\pi(s)$ which maximizes the expected accumulated reward when transitioning through the system. \newtext{Note that this definition entails that if the system is provided with different reward functions, the obtained policies will be different.} However, in most real problems, determining the probabilities or rewards that define a specific \mdp is not an easy task, being in most of the cases unknown, or estimated from noisy observations.

Reinforcement Learning (\rl) is an area of Machine Learning which can tackle the problem of finding an optimal policy $\pi(s)$ for a \mdp where the probabilities or rewards functions are unknown. Specifically, Q-Learning has been proved as a valid \rl algorithm able to find the optimal policy following dynamic programming techniques. Q-Learning is a model-free algorithm which can handle problems with unknown stochastic transitions based on an infinite-time exploration of the transitions between states when different actions are applied based on a partially-random policy of choosing actions. However, due to the infinite-nature of the formulation of the algorithm, the optimality of the obtained policy $\pi'(s)$ will be ultimately based on the time the algorithm has been exploring the system and, in essence, the distribution of the number of times each pair {\em state/action} has been explored. In real-time problems, where the definition of the state comes from real measurements of the environment, this exploration time depends on the frequency the system provides the different metrics, producing long time training sessions. Even more, if the final goal is to train the system several times to obtain different behaviours (changing the reward functions each time), the total training time can extend to unacceptable times.
In the following, we address both questions, namely: {\em (i)} how to define the system to obtain different learned behaviours, and {\em (ii)} how to boost the learning time providing the transition probabilities to the Q-Learning system.

\subsection{Learning different policies}
\label{subsec:learning_different_policies}
Traditionally, training a Q-Learning system is a trial and error process in which the designer of the experiment modifies and tunes the state and reward function definitions until the obtained policy meets the desired behaviour. 
Here, we propose efficient mechanisms to define the state and reward functions to obtain different policies with different behaviours with minimum effort.

\textit{State definition:} Instead of having a unique and complete definition of a state, we propose to divide the states space into different independent sub-spaces (i.e. the metrics required to build one subspace are not used to create any other subspace). This state definition allows us to have a more fine-grained control on the behaviour of the system, and provides a direct way to check how a change in the definition of the problem affects to each subspace independently from the others:
$$s=(s_1, \ldots, s_n), \text{ with } s\in\mathcal{S}=\mathcal{S}_1\times\ldots\times\mathcal{S}_n,\, s_i\in\mathcal{S}_i$$

\textit{Reward definition:} Similarly to the state definition, instead of having a unique definition of reward function, we propose the use of multiple sub-reward functions, each one valuing the goodness of each sub-state; they can be combined into an unique reward through the use of different coefficients:
$$R(s) = R(s_1,\ldots,s_n) = \alpha_1*R_1(s_1) + \ldots + \alpha_n*R_n(s_n), \text{ with } s_i\in\mathcal{S}_i,\,\alpha_i\in\mathbb{R}$$
Given this definition of a reward function, training the system to obtain different policies boils down into tuning each sub-reward to obtain the desired behaviour\newtext{, by means of a tuning-and-test cycle as described in Section~\ref{subsec:methodology}}. This tuning process can be carried out in two different ways based on the desired behaviour:

\textit{(a) Modifying the coefficients:} Assuming all sub-reward functions have the same range (i.e., all functions produce values in the same interval), each coefficient $\alpha_i$ represents the importance of each sub-state in the problem. Modifying these coefficients allows us to give more or less importance to each sub-state, and consequently to the metrics used to build it.
    
\textit{(b) Modifying the reward functions:} Each sub-reward function represents how the system will behave respectively to each sub-state (and, therefore, to each metric used to define each sub-state). The goal of modifying the definition of a sub-reward is not to change the importance given to a set of metrics as before, but instead to modify the behaviour of the system w.r.t. to these metrics. For instance, changing one sub-reward function can imply modifying the behaviour from maximizing certain metric to minimize it.

\subsection{Boosting learning time}
\label{subsec:boosting_learning_time}
Classical Q-Learning formulations are based on a table combining all the actions and states, and representing the expected rewards obtained at each pair. This table, originally empty, is updated at the same time the system explores the different transitions. 
Implicitly, at the same time the table is updated, the system is learning the unknown probabilities between states and actions ($\mathcal{P}$).

Although at first glance the bottleneck of the algorithm seems to be the number of the iterations the system needs to perform in order to explore all the transitions enough number of times, in real-world problems it is limited by the frequency at which the actions can be applied and the metrics needed to build the states can be measured. For example, in a situation in which the system is applied to a video encoding process at 24 frames per second, and a transition occurs between frames, the speed of the exploration is limited to 24 transitions per second.
In the case when multiple policies are required, and consequently, one training session per policy is needed, this limitation in the speed of the algorithm can make the problem unfeasible. 

However, if the probability set between states ($\mathcal{P}$) is known a priori, the algorithm does not need to wait for actual readings of the required 
environmental metrics to determine movements between states; 
contrary, it can simulate the transitions based on the information provided by $\mathcal{P}$. Following this idea, we propose an {\it offline learning process} which dramatically reduces the amount of time needed to obtain each policy: (1) First, all combinations between states and actions are explored enough number of times to build a transition table ($\mathcal{P'}$) that stores the probabilities of moving from one state to the others when applying a specific action. (2) Once $\mathcal{P'}$ has been built, each training process can proceed with the classical Q-Learning formulation. However the state is determined based on the information provided by $\mathcal{P'}$, not by observation. 
The process of building $\mathcal{P'}$ can be carried out independently of the learning process, or can be extracted from one previous training process storing explicitly the probabilities at the same time the classical Q-Learning algorithm explores the different transitions of the system. %
\begin{newtext_env}%
The obtained $\mathcal{P'}$ table will be valid for all the subsequent learning processes, unless the transitions between states change due to external factors (for example, processor operating frequency can be altered by changes in the temperature), or there are changes in the definition of states or actions. Note that $\mathcal{P'}$ is used only in the learning process, but once the system has finished to learn, the states and transitions are obtained directly from measurements of the system, and not from $\mathcal{P'}$.
\end{newtext_env}

%% file: s3-policies_results.tex
\section{Policy extraction for heterogeneous QoS}
\label{sec:policies_results}

\subsection{A case study: video transcoding with heterogeneous QoS requirements}

In previous works~\cite{DATE19}, authors have showed that a multi-agent Q-Learning approach can be applied to a real-world HEVC (High Efficiency Video Coding)~\cite{HEVC} encoding application to obtain simultaneous real-time encoding processes (that is, with a tight lower bound in throughput of 24 frames per second --FPS--), 
and meeting at the same time constraints in terms of quality, bandwidth, and power consumption. 
In that work, number of threads ($N_{th}$), core operating frequency 
({\em  freq}) and the value of the QP encoding parameter~\cite{HEVC}
(higher QP values imply lower encoding quality) were
considered as dynamic knobs, and their values were periodically tuned while throughput, 
power consumption and quality were constantly monitored. 
However, in that scenario, a {\em single} common policy was applied to all the videos 
concurrently processed by the system. We report next the necessary changes to consider
multiple heterogeneous QoS demands, and hence multiple policies.

Consider a similar formulation of the problem in which each state is decomposed into three different sub-states: {\em throughput} (in terms of FPS), {\em quality} (PSNR, measured in dB) and {\em power consumption} (W). Consider also a reward function composed by three different sub-rewards functions, each one associated to a specific sub-state, and its coefficients as described before. In this scenario, and assuming the state definition stays constant, this simple formulation offers 9 different dimensions (3 reward definitions $\times$ 3 coefficient values) the designer of the experiment can tune to obtain different policies.

\begin{figure}
\centering
    \begin{minipage}{0.44\textwidth}
    \includegraphics[width=\textwidth]{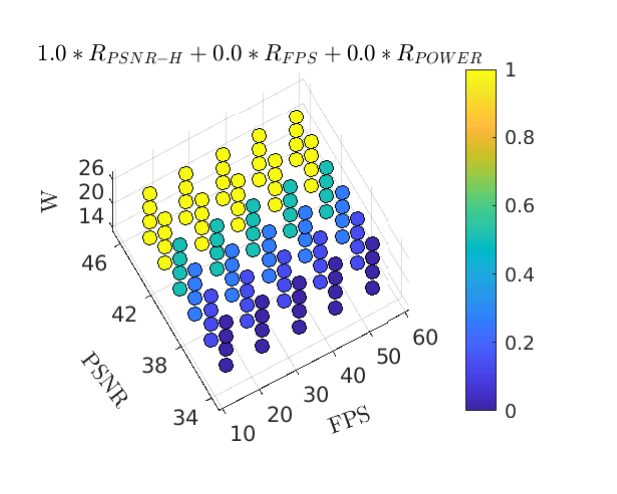}\\
    \includegraphics[width=\textwidth]{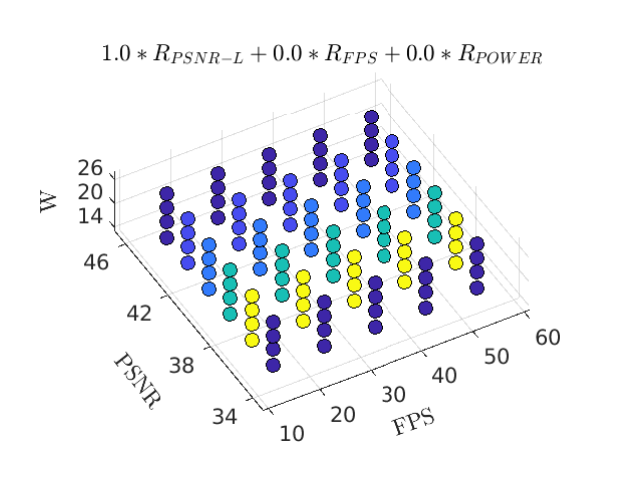}\\
    \includegraphics[width=\textwidth]{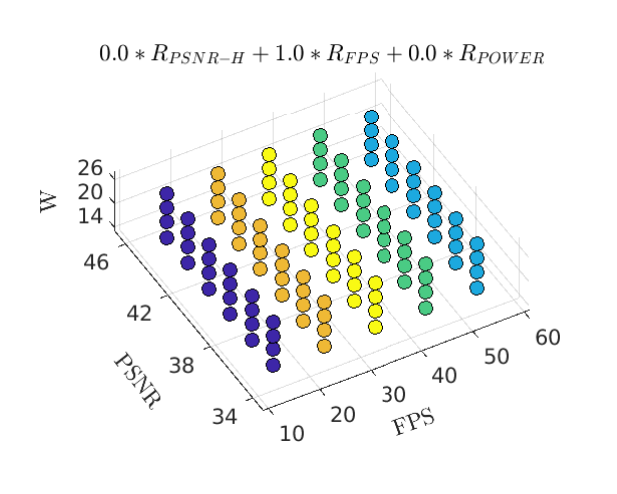}\\
    \end{minipage}
    \begin{minipage}{0.44\textwidth}
    \includegraphics[width=\textwidth]{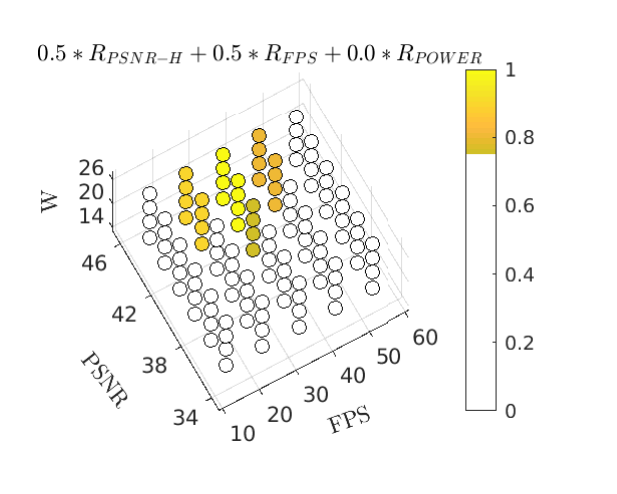}\\
    \includegraphics[width=\textwidth]{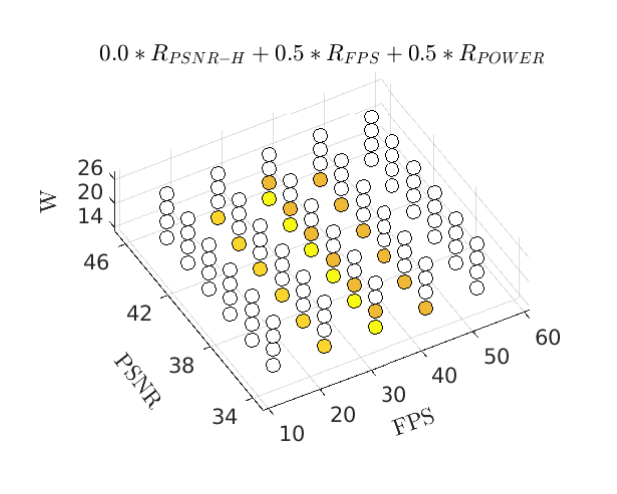}\\
    \includegraphics[width=\textwidth]{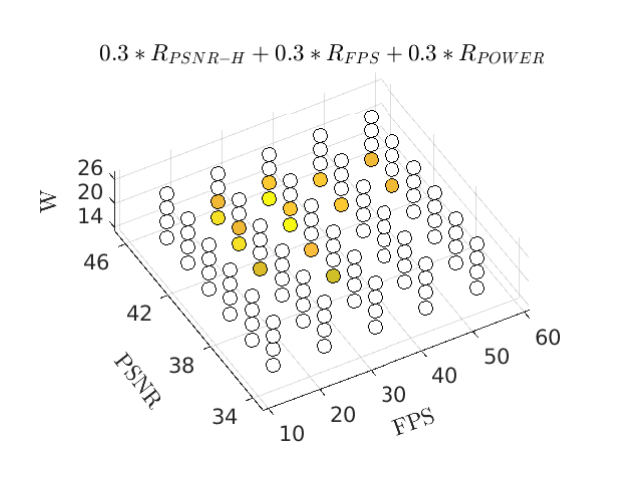}\\
    \end{minipage}
    \caption{Rewards obtained in the different states for different combinations of sub-reward definitions (left) and coefficients (right)}
    \label{fig:rewardsExploration}
\end{figure}

Figure~\ref{fig:rewardsExploration} shows different combinations of sub-rewards and coefficient values. Each dot in the plot represents a state value; its color represents the reward given to that state (higher is better). 
On the left, three different examples of sub-reward functions are shown: (1) \RpsnrH which gives maximum reward to the states with maximum quality; (2)  \RpsnrL which minimizes quality, but ensures a minimum quality (giving a reward of 0 to the states below the threshold); and (3) \Rfps which aims at obtaining real-time encoding processes giving no reward to the states with throughput below 24 FPS, maximum reward to the states between 30 and 40 FPS, and a decreasing reward to the states above.
Observe how modifications in the sub-reward functions \newtext{alter which are the better solutions, or goal states (yellow points in the figure)}. This fact can be clearly seen in the first two plots: while the first one maximizes quality, 
changing the definition in the other one \newtext{turns the states with minimum quality, but above the constraint (36dB in the plot), into the goal states}. 
On the right, different combinations of the same sub-reward functions with different coefficients are shown. For the sake of clarity, only the states with reward $\ge 0.75$ are colored. Here, \Rpower minimizes the power consumption. Observe how small changes in the way functions are combined dramatically alter the goal states of the system. For example, results vary from maximizing quality and ensuring real-time throughput on the top, to restricting the goal space to those with minimum power, meeting real time requirements in the middle\newtext{, and a combination of both behaviours on the bottom}. This figure shows how our proposal offers a plethora of different combinations the designer of the experiment can test, and how a huge number of different policies can be obtained following the methodology described next. 

\subsection{\newtext{Methodology to extract multiple policies}}
\label{subsec:methodology}

Consider a scenario in which a video provider needs to attend multiple video encoding requests from different users with different requirements: \regular users which require a minimum of quality, and \premium users which need encoded videos with maximum quality, with real-time results for both. In this scenario, two different policies are desirable: one which maximizes quality for premium users (\policyP), and another which guarantees a minimum quality for regular users (\policyR). Additionally, when the server load increases, it would be desirable to apply different policies to minimize the resources used by each user, satisfying a minimum quality to each type of user, and still maintaining real-time throughput (24 FPS) (policies \policyPm and \policyRm).

\begin{newtext_env}
Following the previous ideas, we propose a simple and concise methodology to generate the different policies spending a reasonable amount of time and effort in the process. In a first step, the definition of the different states and rewards is done. Each substate needs to be discretized based on expert knowledge of the problem, while the reward functions are defined based on the goal the policy has to achieve. In this step, coefficients are set together with each reward function. In a second step, a detailed simulation of the reward functions for the different states is performed to check if the goal states are the right ones (similar to the space exploration shown in Figure~\ref{fig:rewardsExploration}). If the states with higher reward are not the desired ones, the previous step is repeated until the reward functions and coefficients are properly
tuned. In the last step, the system is trained with the previous rewards and coefficients. If the obtained results are not the desirable ones, the process is repeated until the states, rewards and coefficients are tuned. Although this process can traditionally last for an unacceptable amount of time, note that leveraging the construction of $\mathcal{P'}$, as
described in Section~\ref{subsec:boosting_learning_time}, makes this process relatively fast and easy.
\end{newtext_env}

\begin{figure}
\captionsetup[subfigure]{labelformat=empty,position=above,justification=centering}
    \centering
    \subfloat[\RpsnrL]{
        \includegraphics[width=0.30\textwidth]{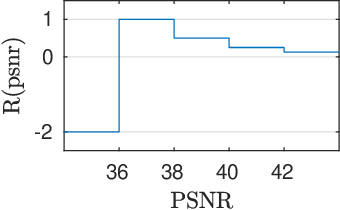}}%
    \subfloat[\RpsnrH]{
        \includegraphics[width=0.30\textwidth]{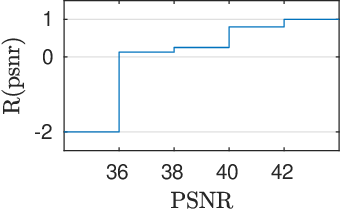}}%
    \subfloat[\RpsnrM]{
        \includegraphics[width=0.30\textwidth]{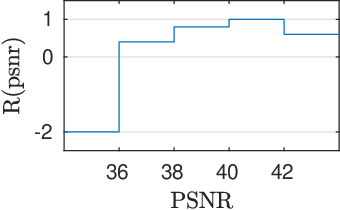}}

    \subfloat[\Rfps]{
        \includegraphics[width=0.30\textwidth]{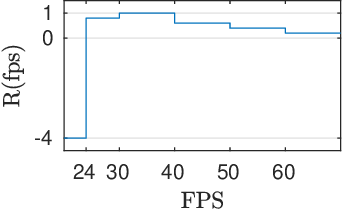}}\quad
    \subfloat[\Rpower]{
        \includegraphics[width=0.30\textwidth]{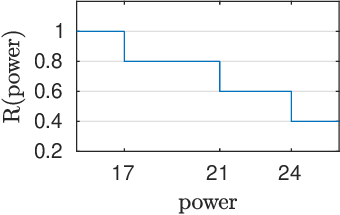}}
    \caption{Sub-reward functions used for the different policies: three sub-rewards for different levels of quality (above), and the functions for real-time encoding and power (below)}
    \label{fig:reward_functions}
\end{figure}

Back in our scenario, the sub-rewards shown in Figure~\ref{fig:reward_functions} where used
to generate the policies: three different reward functions which provide three different levels of quality (\RpsnrL, \RpsnrM, and \RpsnrH), a reward function which guarantees real-time encoding (\Rfps), and a reward function which minimizes power (\Rpower). 
To guarantee a minimum of quality, the functions shown in Figure~\ref{fig:rewardsExploration} were slightly modified to give a negative reward  to those states below the threshold (\PSNR $<36$). Similarly, the definition of \Rfps gives a negative reward to the states below real-time; in this case, a lower reward is given to ensure that videos are always encoded in real-time, not having processes with high quality but throughput below 24 FPS. %
The maximum reward of \Rfps is not given to 24 FPS, but to the next state, due to the fact that, being close to 24 FPS will produce a higher amount of frames being encoded below the threshold due to the variability on the content between frames.
To estimate the power consumption of each application, the model \mbox{$P=\text{nth}*(\alpha*\text{freq}^2+\beta)+\gamma$} was used, setting experimentally the parameters for our specific platform, obtaining a root mean squared error of $0.97W$ and a maximum error of $2.4W$, which are negligible in our machine with a maximum energy consumption of $125W$.

\begin{figure}
\small
\begin{align*}
\pi^{R}:\,& 0.7*R_{PSNR-L} + 0.1 * R_{POWER} + 0.5 * R_{FPS} \\
\pi^{P}:\,& 0.7*R_{PSNR-H} + 0.0 * R_{POWER} + 0.5 * R_{FPS} \\
\pi_{R}:\,& 0.7*R_{PSNR-L} + 0.5 * R_{POWER} + 0.5 * R_{FPS} \\
\pi_{P}:\,& 0.7*R_{PSNR-M} + 0.5 * R_{POWER} + 0.5 * R_{FPS} 
\end{align*}
\caption{Reward functions defined for each policy}
\label{eq:policies}
\end{figure}

We identify \policyR as our base policy, used to tune and polish the actions and state definitions, at the same time the transition table $\mathcal{P'}$ is recorded. Once this policy is created, other policies can be easily derived in a reasonable time \newtext{by means of the described methodology}. Our policies, detailed in Figure~\ref{eq:policies}, were defined based on the following ideas:
\begin{itemize}
    \item For defining \policyR, a combination of a function which minimizes \PSNR but ensures a minimum quality (\RpsnrL), and a function which ensures real-time encoding (\Rfps) were used. Additionally, a reward to minimize power consumption was incorporated with a small coefficient (\Rpower).
    \item The major difference between \policyR and \policyP is the reward function used to evaluate the quality. The former minimizes quality, and the latter maximizes it (\RpsnrH). To achieve high quality videos without violations of throughput, the reward which minimizes power was totally removed. As real-time
    encoding is mandatory in both cases, this term was not modified.
    \item If resource minimization is desired for a regular user, \policyRm achieves that goal by increasing the coefficient of the function which minimizes the power consumption respectively to its counterpart \policyR (from $0.1$ to $0.5$).
    \item The design decisions to create \policyPm are similar to the ones used to create \policyRm, but in this case, because obtaining high quality videos is a resource-hungry process, the sub-reward function associated with PSNR was changed to obtain still high quality videos, but lower quality than in \policyP (\RpsnrM).
\end{itemize}

\begin{table}
\begin{newtext_env}
    \centering
    \begin{tabular}{lcc}\toprule
         & N. frames & Learning Time  \\\midrule
        \sc MonoAgent  & $n\times3000$ & $n\times17h$\\
        \sc MultiAgent & $n\times500$  & $n\times3h$ \\
        \sc This work  & $500 + (n-1)\times500$ & $3h + (n-1)\times1min$\\\bottomrule
    \end{tabular}
    \caption{Learning time to obtain $n$ different policies by different approaches. The Mono- and Multi-agent approaches are those described in~\cite{DATE19}. Learning time has been calculated assuming a learning rate $\approx24fps$}
    \label{tab:learning_time}
\end{newtext_env}
\end{table}

\newtext{Following the ideas described above to boost learning time, once the first policy (\policyR) was defined, the training time to obtain the remaining policies was reduced from days to hours, as shown in Table~\ref{tab:learning_time} when compared against other traditional approaches described in~\cite{DATE19,Iranfar18}.
Note that, in this example, all learning times were extracted using the same machine and setup; also, observe that
our approach inherits the advantages of the Multi-agent implementation (in terms of a reduction in the number of necessary frames
to converge from 3000 to 500 compared with a Mono-agent approach), and adds additional gains as the number of desired policies is increased. In this case, adding a
new policy is translated into roughly one extra minute of computing time. In the case of the traditional Mono- or Multi-agent approaches, each new policy 
would require a complete learning process, adding 17h and 3h per policy, respectively.
}

\subsection{Experimental setup and obtained results}
The described framework and techniques have been implemented in a real server using MAMUT, a centralized resource manager described in~\cite{DATE19}. The HEVC encoder used was Kvazaar~\cite{kvazaar}, an open-source video encoder able to achieve real-time encoding processes. High Definition/1080p ($1080\times720$pixels) video sequences were used in all the experiments, mainly extracted from the JCT-VC benchmark~\cite{jvct-videoset}, using three sequences to train the system, and four different sequences to run the experiments and report results (identified as v1\ldots v4). In the following, the reported results correspond to average values obtained after 5 repetitions of each experiment.

\begin{table}
    \caption{Average knob values learnt by the system for Regular (top) and Premium (bottom) users with and without resource minimization, and output metrics for the different videos used to validate the system}
    \label{tab:resources_regUser}

    \centering
    \begin{tabular}{lllllllllllllll}\toprule
    &\multicolumn{2}{c}{$N_{th}$} && \multicolumn{2}{c}{Freq} && \multicolumn{2}{c}{QP} && \multicolumn{2}{c}{{\sc Quality}} && \multicolumn{2}{c}{$\Delta$} \\\cline{2-3}\cline{5-6}\cline{8-9}\cline{11-12}\cline{14-15}
        &\policyR & \policyRm && \policyR & \policyRm &&  \policyR &  \policyRm &&  \policyR &  \policyRm && \policyR & \policyRm\\\midrule
    v1  &4.2 & 3.3 && 1.6 & 1.7 && 35.2 & 36.0 && 39.4 & 39.1 && 6.0 & 2.4 \\
    v2  &3.1 & 2.9 && 1.4 & 1.5 && 37.0 & 36.9 && 37.8 & 37.8 && 0.4 & 0.4 \\
    v3  &3.1 & 2.6 && 1.3 & 1.5 && 37.0 & 37.0 && 38.0 & 38.0 && 0.3 & 0.3 \\
    v4  &3.3 & 2.9 && 1.3 & 1.4 && 37.0 & 37.0 && 37.1 & 37.1 && 0.4 & 0.4 \\\midrule
   avg. &3.4 & 2.9 && 1.4 & 1.5 && 36.6 & 36.7 && 38.1 & 38.0 && 1.8 & 0.9 \\\bottomrule
    \end{tabular}\\
    \quad\\
    \quad\\
    \begin{tabular}{lllllllllllllll}\toprule
    &\multicolumn{2}{c}{$N_{th}$} && \multicolumn{2}{c}{Freq} && \multicolumn{2}{c}{QP} && \multicolumn{2}{c}{{\sc Quality}} && \multicolumn{2}{c}{$\Delta$} \\\cline{2-3}\cline{5-6}\cline{8-9}\cline{11-12}\cline{14-15}
        & \policyP & \policyPm && \policyP & \policyPm && \policyP  & \policyPm  && \policyP  & \policyPm  && \policyP & \policyPm \\\midrule
    v1  & 4.6 & 3.6 && 1.8 & 1.6 && 24.9 & 33.4 && 43.7 & 40.2 && 2.0 & 6.3 \\
    v2  & 4.2 & 3.3 && 1.7 & 1.4 && 24.9 & 32.2 && 43.2 & 40.2 && 2.3 & 4.2 \\
    v3  & 4.2 & 3.1 && 1.6 & 1.3 && 23.7 & 33.0 && 43.4 & 39.9 && 0.6 & 0.7 \\
    v4  & 4.7 & 3.1 && 1.7 & 1.3 && 22.1 & 33.2 && 43.2 & 39.0 && 0.7 & 0.5 \\\midrule
   avg. & 4.4 & 3.3 && 1.7 & 1.4 && 23.9 & 33.0 && 43.4 & 39.8 && 1.4 & 2.9 \\\bottomrule
    \end{tabular}
\end{table}

Table~\ref{tab:resources_regUser} reports the behaviour of each described policy applied to each video, showing the average knob values set for number of threads ($N_{th}$), frequency (in GHz) and QP, and the output metrics recorded: quality (\PSNR measured in dB), and real time throughput violations (measured as the percentage of time the video has been encoded below 24 FPS \mbox{(-$\Delta$-)}). First, observe how changes in one reward function can produce opposite behaviours. Consider, for example, policies \policyR and \policyP: by modifying exclusively the reward function in charge of quality, the obtained \PSNR changes drastically (38.1 to 43.4 dB, respectively). 
On the contrary, observe how modifications in the coefficients without altering the reward functions can keep the global behaviour of the system intact, but modify the internal actions chosen by the system. Comparing policies \policyR and \policyRm, both achieve comparable quality levels (38.1 vs 38.0 dB), but the number of threads used when the policy \policyRm is acting decreases down to 1 thread in average (in the case of v1 sequence) with respect to \policyR. Regarding policies \policyP and \policyPm, observe how the impact in the number of threads is the same as in the other policies, but the change in the reward which affects quality produces slightly lower \PSNR.

\begin{figure}
    \centering
    \subfloat[Regular users]{%
        \includegraphics[trim=25 20 35 30,clip,width=0.48\textwidth]{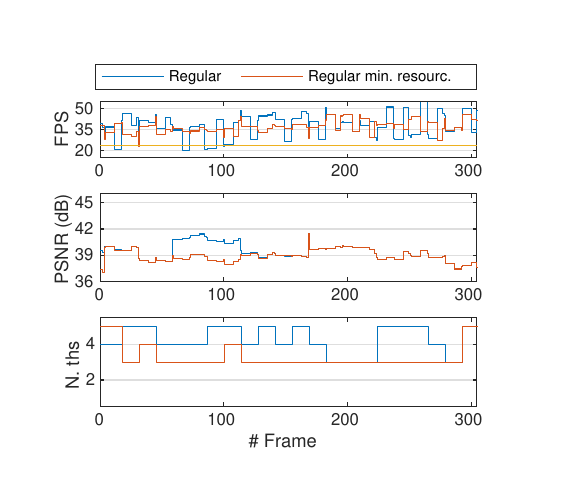}
    }
    \hfill
    \subfloat[Premium users]{%
        \includegraphics[trim=25 20 35 30,clip,width=0.48\textwidth]{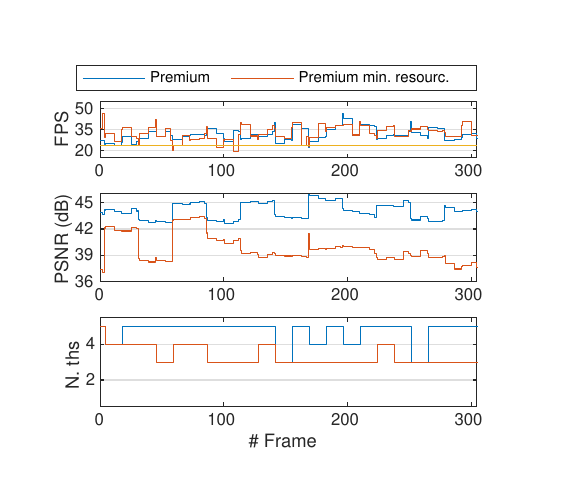}
    }
    \caption{System behaviour timelines and metrics obtained when encoding the v1 sequence with all different policies for Regular users (left) and Premium users (right). The yellow line indicates real-time encoding (24 FPS)}
    \label{fig:timeline_policies}
\end{figure}

Figure~\ref{fig:timeline_policies} shows a timeline of the encoding process of the sequence v1 with the different policies. For clarity reasons, only the changes in number of threads is shown, but dynamic adaptation of frequency and QP values are also performed in the process. In general terms, observe how in the left, in the case of regular users, both policies \policyR and \policyRm obtain similar (and low) quality but the number of threads used in each one changes drastically. On the right, both policies obtain higher quality results, but there is a clear difference in quality and resource usage between them as desired in our formulation.

%% file: s4-heuristic_design.tex
\section{Combining multiple policies into a global system}
\label{sec:heuristic_design}

In the previous scenario, if the number of simultaneous requests is large enough, it may be the case that attending all incoming request simultaneously exhausts the available resources in the server, not being able to encode the different sequences on real time. In this situation, it is common to enqueue the requests, and attend them in order as the previous videos finish and enough resources are available. Of course, the time a request is hold in the queue does not depend only on the number of videos being encoded, but also on the type of user who made the request. 
To reduce the waiting time of each client on the queue, in this section we propose a 3-tier heuristic that, based on the previous obtained policies, is able to reduce the delays thanks to choosing the appropriate policy to apply to each client at each moment. This heuristic is a simple example of how a simple approach can benefit from having different policies to apply, obtaining better results than other static approaches. \newtext{Moreover, note that the methodology described in the previous sections is also valid for any other approach based on multiple simultaneous policies.}

\subsection{A 3-tier heuristic leveraging KaaS (\heupol)}

The proposed heuristic is based on a state machine with three different states, namely: 
{\em (1)} S0, the initial state. This state is active when there are enough resources for all the clients, so that resource usage reduction is not needed. In this state, policies \policyR and \policyP are applied to regular and premium users respectively. 
{\em (2)} If there are not enough resources to encode all the requests, state S1 will try to reduce the resource consumption of the regular users applying policy \policyRm. In this state, premium users are still allowed to use as many resources as they require (policy \policyP). 
{\em (3)} When there are not enough resources for the incoming request in the state S1, and only if the incoming request arrives from a premium user, the heuristic will move to state S2, minimizing the resources for both kind of users, regular and premium. In this state, policies \policyRm and \policyPm are used. Algorithm~\ref{alg:heuristic} shows a detailed pseudocode of the heuristic.


\begin{algorithm}[t]
\caption{\heupol\label{alg:heuristic}}
\Begin{
cl = queue.first\; 
newState = currentState\;

\uIf{(canRunClient(currentState, cl, NReg, NPrem))}
{newState = currentState\;}
\uElseIf{(currentState == S0 \textbf{and} canRunClient(S1, cl, NReg, NPrem))}
{newState = S1\tcp*{Try to move to S1}}
\uElseIf{(cl.type==PREMIUM \textbf{and} canRunClient(S2, cl, NReg, NPrem))}
{newState = S2\tcp*{Try to move to S2}}
\Else{
 queue.insert\_front(cl)\tcp*{Not enough resources}
\Return false\;
}
currentState = newState\tcp*{Update state}
\leIf{(client.type==REGULAR)}{nReg++}{nPrem++}
\Return runClient(cl)\tcp*{Start encoding request}
}
\end{algorithm}

To determine if a new video can be encoded in a given state, or if the system needs to move to another state in order to free resources, a prediction of the number of cores in use is performed. 
\heupol stores in an internal table the average number of cores used by each policy with the training videos, as an estimation value of the cores used by future videos. 
Knowing at each moment the policies that are in use (i.e., the current state), the number of regular and premium users being attended, and the incoming user type, a prediction of the total number of cores in use is calculated. If the amount of predicted cores is lower or equal to the number of physical cores of the machine, the request is attended and the video encoding can commence.  Else, the heuristic will try to move to the next state. If there is not enough room for the user in any state, the request is pushed in front of the queue again. Inserting the client in the front (instead of enqueueing it again in the back) allows to attend the users in the arrival order.
When a video finishes to be encoded and some resources are free, the heuristic checks if it can move to a previous state that does not minimize the resource usage and provides greater quality.

%% file: s5-heuristic_results.tex
\section{Experimental results}
\label{sec:heuristic_results}


For the sake of realism, we assume that several different videos from different users arriving over time need to be served simultaneously, minimizing the waiting time of each client and meeting requirements in quality (based on the type of user) and throughput ($\ge$24 FPS).
Each experiment is determined by the arrival rate (5s, 10s and 15s), and the percentage of premium users (0\%, 25\%, 50\%, 75\% and 100\%). Each experiment comprises 10 sequences to be encoded, randomly selected, with a duration of 2500 frames each ($\approx 100$ seconds at 24 FPS). To obtain reliable data, each configuration of frequency and premium/total users relation was explored through 5 different combinations of 10 videos, and each combination was run 3 times, reporting average values. Although three different arrival frequencies where explored in the experiments (5s, 10s and 15s), only the results of 10s are shown in the following, obtaining similar and comparable measurements at the other frequencies.

For the sake of comparison, we have implemented two additional alternatives to the heuristic approach (\heupol) presented in Section~\ref{sec:heuristic_design}. Both alternatives implement a static decision making process, choosing the policy to serve each video based on the kind of user, and not on the environment. The policy utilized does not change during the whole encoding process:

\begin{enumerate}

\item \onepol: This strategy corresponds to a traditional Q-Learning system using only one policy (\policyR) to encode all the sequences as an extreme case, trading off quality for throughput, as implemented by MAMUT~\cite{DATE19}.

\item \twopol: Due to the existence of different user types with different requirements, we have implemented a Q-Learning system using two different tables, depending on the user type attended (\policyR and \policyP). This policy is the opposite to the previous one: it offers maximum quality to each type of user, without considering decreasing quality to serve more users simultaneously.
\end{enumerate}
In both cases, the algorithm to determine whether a video can be encoded or needs to wait on the queue is the same as that used in \heupol.

\subsection{Comparative performance discussion}

\begin{figure}
    \centering
    \includegraphics[width=0.9\textwidth]{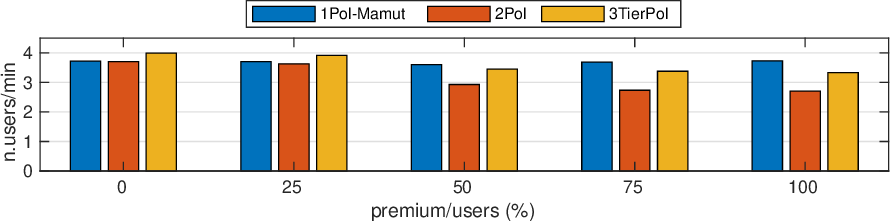}
    \includegraphics[trim={0 0 0 15},clip,width=0.9\textwidth]{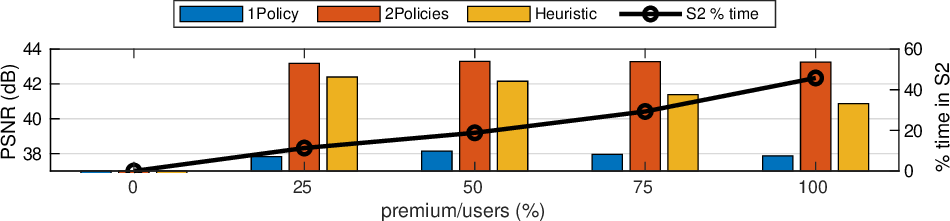}
    \caption{Users per minute attended by each approach with users requests arriving every 10 seconds (top), and quality obtained by each approach when encoding videos from premium users (bars), and percentage of time \heupol is in the S2 state (line)}
    \label{fig:heuristic_results}
\end{figure}

The plot on the top of Figure~\ref{fig:heuristic_results} shows the amount of users attended per minute (on average) based on the number of premium user requests for the three explored approaches. 
Depending on the number of premium users, two different behaviours
are observed: when the amount of premium users is below $50\%$, and when the amount is greater or equal. On the first group, \heupol outperforms the other strategies, and it is able to process more users per minute reducing slightly the quality obtained ($0.13$ dB for regular and $1.9$ dB for premium users in the worst case), see the bottom plot in Figure~\ref{fig:heuristic_results}.

Diving into details of the behaviour of the second group (premium $\ge 50\%$), we observe how the amount of premium users impacts in the performance of the system: as many premium users are attended, more resources are used to encode these videos, and less resources are available to encode new incoming requests. This behaviour is shown in \twopol and \heupol, but not in \onepol. In the case of 
\onepol, the behaviour is to encode all the videos with the same policy (\policyR), used by the others approaches to encode only regular users. On the one hand, this approach uses less resources to process each premium user (because they are considered as regular), but on the other hand, the quality obtained (lower plot on the Figure) is quite lower than the other approaches, obtaining not admissible levels for premium users (up to 5.4 less dB in PSNR). Second, when comparing \twopol and \heupol, we can observe how the latter is able to serve more videos at the same time (up to 1.24$\times$), reducing slightly the obtained quality (a loss of 2.4 dB in the worst case). Finally, observing the \PSNR obtained in each experiment (lower plot) we can see the internal behaviour of \heupol: as the number of premium users increases, \heupol needs to move to states that reduce the resource usage at the expense of reducing the quality. This can be seen in the plot at the bottom of Figure~\ref{fig:heuristic_results}, showing the percentage of the time \heupol is in the S2 state depending on the number of premium users. Observe how the time \heupol is at state S2 increases with the number of premium users.


%% file: s6-related_work.tex

%% file: s7-conclusions.tex
\section{Conclusions}
\label{sec:conclusions}

Training a Q-Learning system from scratch can be an ardours work. In addition, if different policies with different behaviour for the same system are needed, training becomes an unfeasible task in practice.
To tackle this problems, in this work we have proposed a methodology to obtain new policies with minimum effort, once the definitions have been tuned for a first desired behaviour. To achieve that goal, two different mechanisms were proposed: defining the reward functions to obtain the desired behaviour, and speeding up the learning time of the problem. Together, these techniques make it feasible to explore multiple combinations without much effort and yield dramatic improvements in 
terms of learning time.
We have shown how our proposal can be applied to a real scenario obtaining different policies to encode videos on real-time with different QoS objectives, ensuring a minimum of quality in all cases and varying the resource usage at 
will of the policies designer.

By means of a realistic scenario where multiple requests have to be attended simultaneously, we have shown how a heuristic approach can handle multiple policies at the same time, being able to decide which policy apply to each user at each moment, obtaining improvements up to $1.24\times$ in the number of users attended per unit of time when comparing with less sophisticated approaches. 